\documentclass[prl,twocolumn,showpacs,preprintnumbers,superscriptaddress]{revtex4}
\usepackage{amssymb}
\usepackage{amsmath}
\usepackage{graphic  x}
\usepackage{epsfig}
\usepackage{amsfonts}

\begin{document}

\title{Pinning quantum phase transition of photons in a hollow-core fiber}
\date{\today }

\begin{abstract}
We show that a pinning quantum phase transition for photons could be observed in a hollow-core one-dimensional fiber loaded with a cold atomic gas. Utilizing the strong light confinement in the fiber, a range of different strongly correlated polaritonic and photonic states, corresponding to both strong and weak interactions can be created and probed. The key ingredient is the creation of a tunable effective lattice potential acting on the interacting polaritonic gas which is possible by slightly modulating the atomic density. We analyze the relevant phase diagram corresponding to the realizable Bose-Hubbard (weak) and sine-Gordon (strong) interacting regimes and conclude by describing the measurement process. The latter consists of mapping the stationary excitations to propagating light pulses whose correlations can be efficiently probed once they exit the fiber using available optical technologies
\end{abstract}

\pacs{42.50.-p,71.10.Pm,42.50.Gy}
\author{MingXia Huo}
\affiliation{Centre for Quantum Technologies, National University of Singapore, 2 Science
Drive 3, Singapore 117542}
\author{Dimitris G. Angelakis}
\email{dimitris.angelakis@gmail.org}
\affiliation{Centre for Quantum Technologies, National University of Singapore, 2 Science
Drive 3, Singapore 117542}
\affiliation{Science Department, Technical University of Crete, Chania, Crete, Greece,
73100}
\maketitle

%\author{Leong Chuan Kwek}
%\affiliation{Centre for Quantum Technologies, National University of Singapore, 2 Science
%Drive 3, Singapore 117542}
%\affiliation{National Institute of Education and Institute of Advanced Studies, Nanyang
%Technological University, 1 Nanyang Walk, Singapore 637616}

The Bose-Hubbard (BH) and sine-Gordon (sG) models have been extremely
successful in describing a range of quantum many body effects and especially
quantum phase transitions (QPT)\cite{books}. Cold atoms in optical lattices
have been so far the most famous platform to implement these models, where
the Mott insulator (MI) to Superfluid (SF) QPT for a weakly interacting gas
in a deep lattice potential was observed\cite{Jaksch,Bloch}. More recently,
it was made possible to tune up the interactions between the atoms in the
gas leading to the realization of the sG model and the Pinning QPT\cite%
{Buchler,Haller}

Alternative platforms in the field of quantum simulations of many body
effects involve ions for quantum magnets \cite{ions}, and more recently
photonic lattices for the understanding of in and out of equilibrium quantum
many body effects\cite{SIPS}. The photon based ideas are motivated by
significant advances in the fields of Cavity QED and quantum nonlinear optics%
\cite{reviews} and have initiated a a stream of works in the many body
properties of both closed and lossy cavity arrays \cite{SIPSother}. More
recently, a new direction has appeared in the field of SCPs where
hollow-core optical fibers filled with cold atomic gases were considered\cite%
{hollow-core}. The strong light confinement and the resulting large optical
nonlinearities in the single photon level predicted for similar systems\cite%
{DPS}, motivated new proposals to observe photon crystallization and
photonic spin-charge separation\cite{Chang-Angelakis}.

We will show here for the first time that is possible to impose an effective
\textit{lattice potential} on the strongly interacting polaritonic gas in
the fiber. This opens the possibilities for a large range of Hamiltonians to
be simulated with photons. As first examples will analyze the simulation of
the sG and BH models. We will show that the whole phase diagram of the Mott
to SF transitions for both models can be reproduced including a
corresponding photonic ``pinning transition''. We conclude with a discussion
on the available tunability of the quantum optical parameters for the
observation of the strongly correlated phases. The latter is possible by
releasing the trapped polaritons and measuring the correlation on the
photons emitted at the other end of the fiber using available optical
technology. We note here that although our description focus on the hollow-core case, 
tapered fibers with cold atoms could also considered\cite{Sague07}
%. For this system it has been shown a non-equilibrium state exists where the  stationary dark state polaritons (DSP) obey a nonlinear Schr
%\"{o}dinger equation of free particles.
%More recently, a two component photonic Lieb Liniger model and a Luttinger liquid of photons has also been derived by increasing the
%number of the control fields and the atomic species.

%%%%%%%%%%%%%%%%%%%%%%%%%%%%%%%%%%%%%%%%%%%%%%%%%%%%%%%%%%%%%%%%%%%%%%%%%%
%%%%%

\begin{figure}[tbp]
\includegraphics[bb=0 0 517 390, width=7 cm]{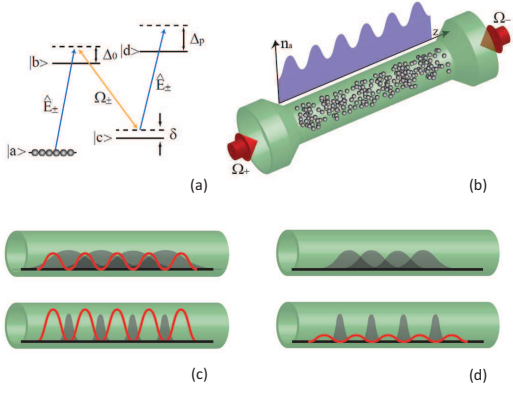}
\caption{(Color online) In (a) and (b) an ensemble of cold atoms with a 4-level
structure is interacting with a pair of classical fields $\Omega_{\pm} $
which create an effective stationary Bragg grating. The photons carried by
the input pulse $E^{+}$ coming in from the left are mapped to stationary
excitations (polaritons) which are trapped in the grating. The strong photon nonlinearity
induced by the 4th level leads to the creation of a strongly interacting gas
 (grey areas in (c) and (d). By modulating the density of
the cold atomic ensemble, an effective lattice potential for polaritons/photons can
be created-shown by red lines in (c) and (d). Tuning to the regime of weak
interactions between polaritons and adjusting the lattice depth, the system
undergoes a BH phase transition from SF (upper one) to MI (lower one) phase
(c). In the opposite regime of strong interactions, the dynamics are
described by the sG model where the addition of an even shallow polaritonic
potential a ``pinning transition'' for polaritons (d) could be observed. As
soon as the desired correlated state is engineered, $\Omega _{-}$
is switched off, and the excitations propagate out of the fiber as
correlated photons. The necessary correlations measurements to probe the
phases of the system can be performed using standard optical technology on
the photons exiting the fiber.}
\label{fig1}
\end{figure}

%%%%%%%%%%%%%%%%%%%%%%%%%%%%%%%%%%%%%%%%%%%%%%%%%%%%%%%%%%%%%%%%%%%%%%%%%%
%%%%%

\textit{Model setup--} The considered atomic level structure is shown in
Fig. \ref{fig1}. The one dimensional cold atomic ensemble is prepared
outside using standard cold atom techniques and is transferred in the fiber
using techniques described in \cite{hollow-core}. The atoms are initially in
the ground state $\left\vert a\right\rangle $
%The density modulation, which as  should be commensurate to the number of quasiparticles,
%be can happen occurs by applying for example
% standing microwave field transfers some atoms
%in $\left\vert a\right\rangle $ to an irrelevant level $\left\vert
%u\right\rangle $ and makes the atoms in $\left\vert a\right\rangle $
%periodic distributed.
and the hollow photonic crystal fiber is injected with a quantum coherent
pulse $\hat{E}_{+}$ from the left side while a pair of classical fields $%
\Omega _{\pm }$ are driving the atomic gas from both two sides (Fig. \ref%
{fig1}(b)). The atomic $\Lambda $ configuration consisting of states $%
\left\vert a,b,c\right\rangle $\ comprises the typical stationary light
set-up \cite{reviews,DPS}. Setting the energy of atomic level $a$\ to be
zero and the atomic levels $b$, $c$, and $d$ to $-\omega _{q}+\Delta _{0}$, $%
-\omega _{cc}+\delta $, and $-\omega _{cc}-\omega _{q}+\Delta _{p}$, the
Hamiltonian describing the four-level atoms, the quantum and classical
fields, and their interaction is written as (in the interaction picture)%
\begin{eqnarray}
H &=&-\int n_{a}dz\{\delta \sigma _{cc}+\Delta _{0}\sigma _{bb}+\Delta
_{p}\sigma _{dd}  \notag \\
&&+\left( g_{ba}\sigma _{ba}+g_{dc}\sigma _{dc}\right) \left( \hat{E}%
_{+}e^{ik_{q}z}+\hat{E}_{-}e^{-ik_{q}z}\right)  \notag \\
&&+\sigma _{bc}\left( \Omega _{+}e^{ik_{c}z}+\Omega _{-}e^{-ik_{c}z}\right)
+h.c.\},  \label{Model}
\end{eqnarray}

where $\sigma _{ij}\equiv \sigma _{ij}\left( z,t\right) =\left\vert
i\right\rangle \left\langle j\right\vert $, $\hat{E}_{\pm }\equiv \hat{E}%
_{\pm }\left( z,t\right) =\sum_{k}a_{k}e^{\pm i\left( k-k_{q}\right)
z}e^{-i\left( \omega _{k}-\omega _{q}\right) t}$, and $\Omega _{\pm }\equiv
\Omega _{\pm }\left( z,t\right) =\sum_{k}fe^{\pm i\left( k-k_{c}\right)
z}e^{-i\left( \omega _{k}-\omega _{c}\right) t}$. $\sigma _{ij}$ are
collective and continuous operators describing the average of $\left\vert
i\right\rangle \left\langle j\right\vert $ over atoms with density $n_{a}$
in a small but macroscopic region around $z$, where $i$, $j$\ go over $a$, $%
b $, $c$, and $d$. The fields $\hat{E}_{\pm }$ couple the ground state $%
\left\vert a\right\rangle $\ to excited state $\left\vert b\right\rangle $\
with a strength given by $g_{ba}$, and $\left\vert c\right\rangle $\ to $%
\left\vert d\right\rangle $\ with a strength $g_{dc}$.\ The metastable state
$\left\vert c\right\rangle $\ and $\left\vert b\right\rangle $\ are coupled
by classical, counter-propagating control fields $\Omega _{\pm }$. The
quantum field and classical field envelopes are slowly varying operators and
$k_{q}$ and $k_{c}$ are the wavevectors corresponding to their central
frequencies $\omega _{q}$\ and $\omega _{c}$\ for $\hat{E}_{\pm }$\ and $%
\Omega _{\pm }$, respectively\cite{DPS}.

The evolution of $\hat{E}_{\pm }$ in the fiber is described by the
Maxwell-Bloch equation as
\begin{equation}
\left( \partial _{t}\pm v\partial _{z}\right) \hat{E}_{\pm }=-iv\Delta
\omega \hat{E}_{\pm }+i\sqrt{2\pi }n_{a}\left( g_{ba}\sigma _{ab,\pm
}+g_{dc}\sigma _{cd,\pm }\right) ,  \label{Bloch}
\end{equation}%
where the slowly varying operators $\sigma _{ab}=\sigma
_{ab,+}e^{ik_{q}z}+\sigma _{ab,-}e^{-ik_{q}z}$ and $\sigma _{cd}=\sigma
_{cd,+}e^{ik_{q}z}+\sigma _{cd,-}e^{-ik_{q}z}$ are introduced. $v$ is the
light speed in the empty waveguide, $v=\omega _{q}/k_{q}$ and $\Delta \omega
$ is the difference between $\omega _{q}$ and $\omega _{c}$

We assume that the atoms are initialized in the\ ground state $\left\vert
a\right\rangle $ and the quantum field is a weak coherent state containing 
roughly ten photons. Following the standard methods for treating slow light
polaritons analyzed in \cite{DPS,Chang-Angelakis}, and setting the
coupling constants $g_{ba}=g_{dc}=g$ for simplicity, we introduce $\Psi _{+}$%
, $\Psi _{-}$ as the forward- and backward-going polaritons. These are the
propagating excitations and are defined as $\Psi _{\pm }=\cos \theta \hat{E}%
_{\pm }-\sin \theta \sqrt{2\pi n_{a}}\sigma _{ca}$ with $\cos \theta =\Omega
_{\pm }/\sqrt{\Omega _{\pm }^{2}+2\pi g^{2}n_{a}}$\ and $\sin \theta =g\sqrt{%
2\pi n_{a}}/\sqrt{\Omega _{\pm }^{2}+2\pi g^{2}n_{a}}$. In the limit $g\sqrt{%
2\pi n_{a}}\gg \Omega _{\pm }$, i.e. $\sin \theta \simeq 1$, the excitations
are mostly in the spin-wave form, $\Psi _{\pm }=g\sqrt{2\pi n_{a}}\hat{E}%
_{\pm }/\Omega _{\pm }$, with a group velocity given by $v_{g}\simeq v\Omega
^{2}/\left( \pi g^{2}n_{a}\right) $. Next we adiabatically eliminate the
fast rotating term from Eq. (\ref{Bloch}) and set $\Psi =\left( \Psi
_{+}+\Psi _{-}\right) /2$ as the stationary combinations. In the limit of a
large optical depth \footnote{%
OD$=n_{a}L\Gamma _{\mathrm{1D}}/\Gamma $ and $\Gamma _{\mathrm{1D}}=4\pi
g^{2}/v$\ with $\Gamma /\Gamma _{\mathrm{1D}}$ the ratio of the total
spontaneous emission rate to spontaneous emission into the waveguide}, the
equations of motion for $\Psi $ read
\begin{equation}
i\partial _{t}\Psi =-\frac{1}{2m}\partial _{z}^{2}\Psi +V\Psi +2\chi \Psi
^{\dagger }\Psi ^{2}.  \label{NLE}
\end{equation}

The effective mass is $m=-\Delta \omega /(2vv_{g})-\Gamma _{\mathrm{1D}%
}n_{a}/(4\Delta _{0}v_{g})$, the potential strength is $V=\Delta \omega
v_{g}/v-\Lambda \Gamma _{\mathrm{1D}}\delta v_{g}n_{a}/(4\Omega ^{2})$ and
the interaction strength between polaritons is $\chi =\Lambda ^{2}\Xi \Gamma
_{\mathrm{1D}}v_{g}/(2\Delta _{p})$ where $\Lambda =\Omega ^{2}/(\Omega
^{2}-\delta \Delta _{0}/2)$ and $\Xi =(\Delta _{p}-\delta /2)/(\Delta
_{p}-\delta )$.
%In the limit where the slow light $\Delta \omega/(2v)\ll \Gamma _{\mathrm{1D}}n_{a}/(4\Delta _{0})$, the effective mass is
%approximately $m=-\Gamma _{\mathrm{1D}}n_{a}/(4\Delta _{0}v_{g})$.

\textit{Adding a periodic potential--} The nonlinear Schr\"{o}dinger
equation (\ref{NLE}) determines the evolution of the trapped polaritonic
field $\Psi \left( z,t\right) $\ as derived from the effective Hamiltonian, $%
H=\int dz\Psi ^{\dagger }\left( -\frac{\hbar ^{2}}{2m}\nabla ^{2}+V\right)
\Psi +\chi \int dz\Psi ^{\dagger }\Psi ^{\dagger }\Psi \Psi .$ To add an
effective polaritonic lattice, we induce a periodic atomic density
distribution by applying an external field such that the atoms in $%
\left\vert a\right\rangle $ are now given by $n_{a}=n_{0}+n_{1}\cos ^{2}(\pi
n_{\mathrm{ph}}z)$. We keep $n_{0}\gg n_{1}$ which means the modulation is
only a perturbation in the atomic density and derive the new Hamiltonian
which reads%
\begin{eqnarray}
H &=&\int dz\Psi ^{\dagger }\left[ -\frac{\hbar ^{2}}{2m}\nabla
^{2}+V_{0}+V_{1}\cos ^{2}(\pi n_{\mathrm{ph}}z)\right] \Psi  \notag \\
&&+\chi \int dz\Psi ^{\dagger }\Psi ^{\dagger }\Psi \Psi ,  \label{LL}
\end{eqnarray}%
where $V_{0}=\Delta \omega v_{g}/v-\Lambda \Gamma _{\mathrm{1D}}\delta
v_{g}n_{0}/(4\Omega ^{2})$\ is the corresponding trapping polaritonic
potential and $V_{1}=-\Lambda \Gamma _{\mathrm{1D}}\delta
v_{g}n_{1}/(4\Omega ^{2})$ the resulting imposed polaritonic lattice depth.
We stress here the dependance of the effective polaritonic lattice on both the slow light parameters (group velocity, trapping
laser detuning and strength), and the modulated atomic density. Finally,
we note that the atomic lattice modulation should be chosen to be
commensurate to the number of the photons in the initial pulse for the
pinning transition to occur\cite{Haller}. This mean the modulation length will
approximately fall in the microwave regime as the numbers of trapped photons
in the initial pulse is of the order of 10 and the fiber is a few cm in
length.
%%%%%%%%%%%%%%%%%%%%%%%%%%%%%%%%%%%%%%%%%%%%%%%%%%%%%%%%%%%%%%%%%%%%%%%%%%
%%%%%

\begin{figure}[tbp]
\includegraphics[bb=0 0 530 220, width=8.0 cm]{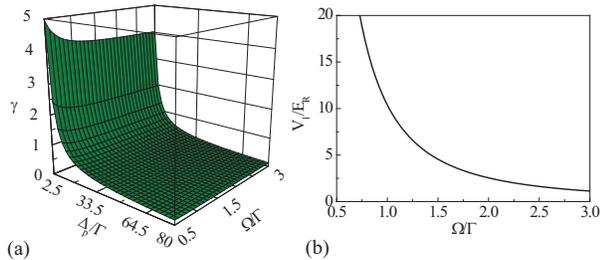}
\caption{(Color online) Plots of the Lieb-Liniger interaction parameter $%
\protect\gamma $ as a function of the one-photon detuning $\Delta
_{p}/\Gamma $ and the Rabi frequency $\Omega /\Gamma $\ of the classical
laser field (a) and the lattice depth $V_{1}/E_{\mathrm{R}}$ as a function
of $\Omega /\Gamma $ (b). The parameters are taken as $n_{a}=10^{7}\mathrm{m}%
^{-1}$, $n_{1}=0.1n_{a}$, $n_{\mathrm{ph}}=10^{3}\mathrm{m}^{-1}$, and $%
\Gamma _{\mathrm{1D}}=0.2\Gamma $, $\Delta _{0}=5\Gamma $, $\protect\delta %
=0.01\Gamma $ with $\Gamma \simeq 20\mathrm{MHz}$ the atomic decay rate. }
\label{fig2}
\end{figure}

%%%%%%%%%%%%%%%%%%%%%%%%%%%%%%%%%%%%%%%%%%%%%%%%%%%%%%%%%%%%%%%%%%%%%%%%%%
\textit{Reaching the relevant correlated regimes--} The success of achieving
a specific strongly correlated polaritonic/photonic state will be characterized by
the feasibility of tuning the Lieb-Liniger ratio of the interaction and
kinetic energies $\gamma $, and the ratio of the depth of the polaritonic
potential to the recoil energy $V_{1}/E_{R}$ to the relevant regimes\cite%
{books,Jaksch}. In our system these two quantities read:
\begin{eqnarray}
\gamma &=&\frac{m\chi }{n_{\mathrm{ph}}}=-\frac{\Lambda ^{2}\Xi }{8}\frac{%
\Gamma _{\mathrm{1D}}^{2}}{\Delta _{0}\Delta _{p}}\frac{n_{0}}{n_{\mathrm{ph}%
}},  \label{plots1a} \\
\frac{V_{1}}{E_{\mathrm{R}}} &=&\frac{\Lambda }{8\pi ^{2}}\frac{\Gamma _{%
\mathrm{1D}}^{2}}{\Omega ^{2}}\frac{\delta }{\Delta _{0}}\frac{n_{0}n_{1}}{%
n_{\mathrm{ph}}^{2}}.  \label{plots1b}
\end{eqnarray}%
Assuming fixed atomic and photonic densities, both quantities can be
controlled by tuning the one-photon detuning $\Delta _{p}/\Gamma $ (shifting
the fourth level $|d\rangle $ to/from to resonance) and by changing the
strength of the control Rabi frequency $\Omega $.
% \footnote{%
%We note that here we are analyzing the repulsive case on a lattice with $%
%\gamma >0$ and mention that the attractive one for the homogenous case in
%the general nonlinear fiber set up, has been studied and leads to
%photonic soliton propagation as in Y. Lai and H. Hauss, \textit{Phys. Rev. A} \textbf{40}
%854 (1989). }. 
In Fig. \ref{fig2} we plot
the achievable regimes for $\gamma $ and $V_{1}/E_{\mathrm{R}}$ as a
function of the $\Delta _{p}/\Gamma $ and $\Omega /\Gamma $\ of the classical
control laser field for realistic parameters. We assume a total atomic decay
rate from the upper level $\Gamma \simeq 20\mathrm{MHz}$, approximately
atomic density is $n_{0}=10^{7}\mathrm{m}^{-1}$ ($10^{5}$\ atoms into a $1%
\mathrm{cm}$ length fiber) and a photonic density of $n_{\mathrm{ph}}=10^{3}%
\mathrm{m}^{-1}$ (the input quantum light pulse containing roughly $10$
photons). For these values, $\gamma $ and $V_{1}/E_{\mathrm{R}}$ can be
tuned in the range $0$ to $5$ and from $0$\ to $20$ respectively, which
allows for both the strong and weak interaction regimes to be realized with
our trapped polaritonic gas. We note that the current state of the art in
the number of atoms loaded in similar hollow-core fibres is smaller by
roughly one order of magnitude or less. However, recent experimental
progress in the field show that our requirements should be satisfied in the
very near future\cite{hollow-core}.  The losses which will
mainly occur due to spontaneous emission from the upper levels, can be
estimated by including the corresponding terms in the Hamiltonian Eq. (\ref%
{LL}). In that case, the effective parameters will acquire an imaginary part
which for the effective mass for example will read $m=-\Delta \omega
/(2vv_{g})-\Gamma _{\mathrm{1D}}n_{a}/(4\Delta _{0}v_{g}+2i\Gamma v_{g})$,
leading to a loss rate $\kappa =\frac{n_{\mathrm{ph}}^{2}v_{g}\Gamma }{%
n_{a}\Gamma _{\mathrm{1D}}}$. These losses will set an upper bound on the timescales for the preparation of the states,
and the probing of the established correlations. For the values under consideration in Fig. 2
and typical slow light velocities $v_{g}$ of 100m/s\cite{DPS}, these translate to lifetimes  of hundreds of micro-seconds.
 The latter are within the reach of current optical measurement technology.
%For the standing microwave field, $%
%\Omega _{\mathrm{M.W.}}e^{i\pi n_{\mathrm{ph}}z}+\Omega _{\mathrm{M.W.}%
%}e^{-i\pi n_{\mathrm{ph}}z}=2\Omega _{\mathrm{M.W.}}\cos (\pi n_{\mathrm{ph}%
%}z)$ and the transfer rate from $\left\vert a\right\rangle $-state to $%
%\left\vert u\right\rangle $-state is $p_{u}\simeq \lbrack 2\Omega _{\mathrm{%
%M.W.}}\cos (\pi n_{\mathrm{ph}}z)]^{2}t^{2}/4=\Omega _{\mathrm{M.W.}%
%}^{2}\cos ^{2}(\pi n_{\mathrm{ph}}z)t^{2}$ \cite{MW}. Switching on the
%microwave field for $t\sim 6\mathrm{ns}$ makes $p_{u}\sim 10\%$\ and
%transfers ten percent atoms from $\left\vert a\right\rangle $\ to $%
%\left\vert u\right\rangle $, corresponding to $n_{1}\sim 0.1n_{0}$. Since
%the wavelength of the microwave field is $\lambda _{\mathrm{M.W.}}=2\mathrm{%
%mm}$, the frequency difference between states $\left\vert a\right\rangle $
%and $\left\vert u\right\rangle $\ is $\omega _{\mathrm{M.W.}}=150\mathrm{GHz}
%$.

\textit{Polaritonic/photonic pinning transition--} We will now discuss the
nature of the many body states generated by the addition of the effective
polaritonic potential and show that a \textquotedblleft pinning transition"
for polaritons can be observed, similar to the one recently experimentally
verified for bosonic atoms in \cite{Haller}. This polaritonic pinning
transition is expected to transform continuously into the BH regime for a
sufficiently deep effective lattices (large $V_{1}/E_{\mathrm{R}}$) and
small interactions $\gamma $. To analyze each relevant phase of the system,
we will make use of the corresponding BH and sG models from many body physics\cite{books}.
 We will also discuss the feasibility to access the whole of the relevant phase diagram for both cases, by simply tuning
the optical parameters in our system.

We first analyze the strong interaction regime $1\leq \gamma \leq 5$, and
for a weak effective potential, $V_{1}/E_{\mathrm{R}}\leq 3$. This regime is
clearly accessible in our photonic system as we show in Fig. 2, by
appropriate tuning of the one photon detuning $\Delta_{p}$ and the control laser
strength $\Omega$. In this case the proper low-energy description of the system
described in Eq. (\ref{LL}), is given by the quantum sG model which reads%
\cite{books}
\begin{equation}
H=\int \frac{dz}{2}\left\{ \frac{\hbar v_{g}}{\pi }\left[ \left( \partial
_{z}\theta \right) ^{2}+\left( \partial _{z}\phi \right) ^{2}\right]
+V_{1}n_{\mathrm{ph}}\cos \left( 4K\theta \right) \right\}.  \label{sG}
\end{equation}%
The first two terms account for the kinetic and interaction energies of
polaritons respectively and $\partial _{z}\theta $\ and $\partial _{z}\phi $
denote the fluctuations of the long-wavelength density and phase fields $%
\theta $\ and $\phi $ \cite{books}. The dimensionless parameter $K=\hbar n_{\mathrm{ph}%
}\pi /\left( mv_{g}\right) $ is known to be related to $\gamma $ as $K\simeq \pi /%
\sqrt{\gamma -\gamma ^{3/2}/\left( 2\pi \right) }$ for $\gamma \leq 10$.

On the other hand, if we tune the system to the weak interaction limit with
small $\gamma \le 1$ and large $V_{1}/E_{\mathrm{R}}$ with $V_{1}/E_{\mathrm{%
R}}\gg 1$, the system is characterized in a good approximation by the BH
model\cite{Jaksch},%
\begin{equation}
H=-J\sum_{i}\left( b_{i}^{\dagger }b_{i+1}+H.c.\right) +\frac{U}{2}%
\sum_{i}n_{i}\left( n_{i}-1\right),  \label{BH}
\end{equation}%
where $J/E_{\mathrm{R}}=4\left( V_{1}/E_{\mathrm{R}}\right) ^{3/4}\exp
\left( -2\sqrt{V_{1}/E_{\mathrm{R}}}\right) /\sqrt{\pi }$, $U/E_{\mathrm{R}}=%
\sqrt{2/\pi ^{3}}\left( V_{1}/E_{\mathrm{R}}\right) ^{1/4}\gamma $.

\begin{figure}[tbp]
\includegraphics[width=5.0 cm]{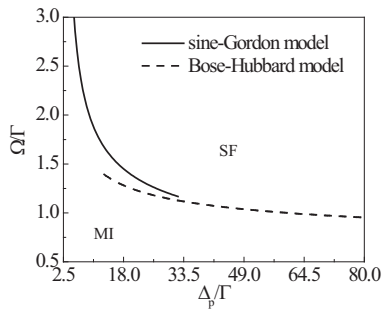}
\caption{The phase diagram for sG model (large $\protect\gamma $ and small $%
V_{1}/E_{\mathrm{R}} $) and BH model (small $\protect\gamma $ and large $%
V_{1}/E_{\mathrm{R}}$). Here $x$-\ and $y$-axis give the range of
tuning parameters $\Delta _{p}$\ and $\Omega $\ needed for two
regimes. The upper region corresponds to SF phase, while the lower
region corresponds to MI phase.The rest of the parameters are as in
Fig. 2.} \label{fig3}
\end{figure}

In Fig. 3, we show that by simply varying $\Delta _{p}/\Gamma $ and $\Omega /\Gamma
$ the whole phase diagram corresponding both to the sG and BH regimes can be
accessed in our system for realistic values of the optical parameters.  We plot the
known phase transition lines corresponding to the sG and BH model
occurring at $V_{1}/E_{\mathrm{R}}=2\pi /\sqrt{\gamma -\gamma ^{3/2}/\left(
2\pi \right) }-4$, and $\left( U/J\right) _{c}=\sqrt{2}\exp \left( 2\sqrt{%
V_{1}/E_{\mathrm{R}}}\right) \gamma /\left[ 4\pi (V_{1}/E_{\mathrm{R}})^{1/2}%
\right] \simeq 3.85$ respectively\cite{books,Bloch,Jaksch}.
In our case, these are probed by adjusting the detuning
and the laser coupling accordingly.
% Generally speaking for larger single photon detunings $\Delta_{p}$ and
%smaller laser strength $\Omega$, the system lies in the weak interacting regime
%described by the BH model, whereas the sG behaviour appears in the opposite
%regime (see Fig. 3). 
The pinning transition is expected to occur for any
value $\Delta _{p}/\Gamma $ less than 20 when $\Omega $ is increased to be
larger than $\Gamma $ which correspond to vanishing lattice and strong interaction regime $\gamma \ge 3.5$.
The BH Mott transition will occur in the opposite weakly interacting regime and deeper lattices. In Fig. 4, for a
specific value of the $\Delta _{p}/\Gamma=50$ corresponding to with  $\gamma \l 1$, we plot he
interaction $U$ and and tunneling strengths as a function of $\Omega
/\Gamma $ to further illustrate this case. We see that transition occurs for
$\Omega /\Gamma \simeq 1.03388$ which corresponds to known
critical point of $\left( U/J\right) _{c}\simeq 3.85$.

%%%%%%%%%%%%%%%%%%%%%%%%%%%%%%%%%%%%%%%%%%%%%%%%%%%%%%%%%%%%%%%%%%%%%%%%%%
%%%%%

\begin{figure}[tbp]
\includegraphics[width=5.0 cm]{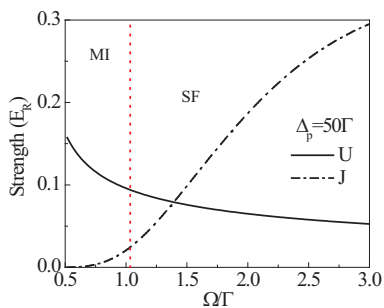}
\caption{ The interaction and tunneling strength as functions of $%
\Omega/\Gamma $ with $\Delta _{p}=50\Gamma $ for the weakly
interacting gas in the BH regime. The red dot line at $\Omega
/\Gamma \simeq 1.03388$\ corresponds to the Mott phase transition
point $\left( U/J\right) _{c}\simeq 3.85$. The rest of the
parameters are as in Fig. 2.} \label{fig4}
\end{figure}

%%%%%%%%%%%%%%%%%%%%%%%%%%%%%%%%%%%%%%%%%%%%%%%%%%%%%%%%%%%%%%%%%%%%%%%%%%

%{\it Losses--} For the small two-photon detuning regime we are considering $\delta \ll \Omega ^{2}/\Delta _{0},\Delta
%_{p} $ the loss rate induced by the atomic decay can be estimated to:
%\begin{equation}
%\kappa _{\mathrm{tot}}=(\frac{1}{2\pi ^{2}}\frac{\Gamma }{\Delta _{0}}+\frac{%
%\delta \Gamma /4}{\Omega ^{2}}\frac{n_{0}}{n_{1}}\frac{V_{1}}{E_{\mathrm{R}}}%
%+\frac{1}{4\pi ^{2}}\frac{\Gamma }{\Delta _{0}}\frac{\Gamma _{\mathrm{1D}%
%}^{2}}{\Delta _{p}^{2}}\frac{n_{0}}{n_{\mathrm{ph}}})E_{\mathrm{R}}.
%\end{equation}%
%In this expression, the first term comes from the kinetic energy term, the
%second one is related to the lattice potential, while the last one is from
%the nonlinear term. Calculations show that by optimizing $\Delta _{0}$, $%
%\kappa _{\mathrm{tot}}\sim 0.1E_{\mathrm{R}}$ around the critical line.

We will now summarize the whole process consisting of preparation, evolution and
detection of the states. First, the atoms are prepared in state $\left\vert
a\right\rangle $ with a slightly modulated atomic density and transferred in the fiber
\footnote{This part depending the exact implementation scheme could also happen the other
way around, i.e., the modulation occurs when the atoms are already in the fiber.}. Next, $\hat{E}_{+}$ is sent in from the left with a
co-propagating control field $\Omega _{+}$ turned on. Once $\hat{E}_{+}$
completely enters the medium, $\Omega _{+}$ is slowly turned off, reversibly
converting $\hat{E}_{+}$ into the pure atomic excitations in the usual slow
light manner\cite{DPS} and the medium is illuminated simultaneously by $\Omega
_{+}$ and $\Omega _{-}$, making $\hat{E}_{+ }$ quasi-stationary and
trapped. The quantities $\gamma $ and $V_{1}/E_{\mathrm{R}}$ are then changed in
time to reach the desired state. Finally one
of the control fields $\Omega _{-}$ is switched off, mapping the polaritons to photons and releasing the excitations.
 At this time, the spatial correlations are mapped into
temporal correlations of outgoing photons, on which measurements are done
using standard quantum optical techniques. By analyzing the resulting
photonic spectra and correlation functions the state of system could be
obtained.

In conclusion we have shown that different strongly correlated states of photons could be created
inside hollow-core fibers interacting with atomic gases with current or near future optical technology. 
The resulting states can be controllably tuned to reproduce sG and BH many body dynamics and
also used to probe the corresponding pinning quantum phase transition predicted by these models.
The various correlated phases can be analyzed by standard optical correlations measurements on the light exiting the fiber.

We would like to acknowledge useful discussion by Profs LC Kwek and KS Huang
in the early stages of this work and the financial support by the National
Research Foundation \& Ministry of Education, Singapore.

\end{document}